\documentclass[manuscript]{aastex631}

\begin{document}

\title{Pickup Ion-Mediated Magnetic Reconnection in the Outer Heliosphere}

\correspondingauthor{Masaru Nakanotani}
\email{mn0052@uah.edu}

\author[0000-0002-7203-0730]{M. Nakanotani}
\affiliation{Center for Space Plasma and Aeronomic Research (CSPAR) and Department of Space Science, The University of Alabama in Huntsville, Huntsville, Alabama 35805, USA}

\author[0000-0002-4642-6192]{G. P. Zank}
\affiliation{Center for Space Plasma and Aeronomic Research (CSPAR) and Department of Space Science, The University of Alabama in Huntsville, Huntsville, Alabama 35805, USA}

\author[0000-0002-4299-0490]{L.-L. Zhao}
\affiliation{Center for Space Plasma and Aeronomic Research (CSPAR) and Department of Space Science, The University of Alabama in Huntsville, Huntsville, Alabama 35805, USA}



\begin{abstract}
     Pickup ions (PUIs) play a crucial role in the heliosphere, contributing to the mediation of large-scale structures such as the distant solar wind, the heliospheric termination shock (HTS), and the heliopause.
     While magnetic reconnection is thought to be a common process in the heliosphere due to the presence of heliospheric current sheets, it is poorly understood how PUIs might affect the evolution of magnetic reconnection.
     Although it is reasonable to suppose that PUIs decrease the reconnection rate since the plasma beta becomes much larger than $1$ when PUIs are included, we show for the first time that such a supposition is invalid and that PUI-induced turbulence, heat conduction, and viscosity can preferentially boost magnetic reconnection in heliospheric current sheets in the distant solar wind.
     This suggests that it is critical to include the effect of the turbulence, heat conduction, and viscosity caused by PUIs to understand the dynamics of magnetic reconnection in the outer heliosphere.
\end{abstract}

\keywords{Classical Novae (251) --- Ultraviolet astronomy(1736) --- History of astronomy(1868) --- Interdisciplinary astronomy(804)}


\section{Introduction} \label{sec:intro}

It has now been established over several decades that pickup ions (PUIs) born through the interaction of interstellar neutral atoms with  the solar wind play a crucial role in shaping the physics and structure of the heliosphere.
The Voyager 2 and New Horizons spacecraft observed the increase of the solar wind temperature in the outer heliosphere, and this has been understood as the result of the dissipation of turbulence driven by newly-born PUIs \citep{williams1995,matthaeus1999, isenberg2003} and modeled well in the upwind and downwind directions \citep{matthaeus1999, zank2018, nakanotani2020}.
The structure of the heliospheric termination shock (HTS) is greatly mediated by PUIs \citep{matsukiyo2014,mostafavi2017,mostafavi2018,kumar2018}.
While PUIs are preferentially reflected at the HTS and gain energy, the core-proton component transmits downstream without a strong heating \citep{zank1996a}.
This may result in a large shock thickness as observed by the Voyager spacecraft, which can be described by a PUI-MHD model with PUI viscosity and heat conduction effects \citep{mostafavi2017}.
Charge-exchange of neutral atoms around the heliopause causes a pseudo-gravitational force that drives a Rayleigh–Taylor-like instability in the nose region \citep{zank1996,zank1999, florinski2005, opher2021} as well as a Kelvin-Helmholtz-like instability \citep{borovikov2008,avinash2014}.
It is possible too that PUIs can modify the nature of low-frequency waves in the distant solar wind, introducing a modified sound speed and moreover, very long-wavelength PUI-modified waves can be driven unstable by PUIs, for which there may be some evidence. \citep{zank2005, fujiki2014}

Although large-scale structures, waves, and discontinuities are mediated by PUIs in the heliosphere, the question of how PUIs affect magnetic reconnection in the heliosphere has not been addressed in any detail.
This is because it is difficult to observationally locate where reconnection occurs, and we had/have only the Voyager and New Horizons spacecraft in the outer heliosphere.
In addition, the Voyager spacecraft magnetometer did not have sufficient resolution to observe reconnection at current sheets, and unfortunately New Horizons does not have instrumentation to measure magnetic fields.
However, because of the possibility that reconnection processes can yield energetic particles in the solar wind \citep{zank2014b,  khabarova2015} and the heliosheath \citep{drake2010, zank2015, nakanotani2021}, it is necessary to understand the effect of PUIs on magnetic reconnection.
An interesting and important question is if PUIs preferentially boost magnetic reconnection, then this may indicate that there are more energetic particles than we currently expect in the outer heliosphere.

It is reasonable to expect that PUIs reduce the reconnection rate since the plasma beta (the ratio of the thermal plasma pressure to the magnetic pressure) becomes much greater than $1$ when the PUI pressure is included.
Supposing that the PUI-plasma system is a single and ideal MHD fluid, the total plasma beta becomes $\beta=\beta_e+\beta_p+\beta_{\rm PUI}$ where the subscripts stand for electron ($e$), and proton ($p$), respectively.
Since the effective PUI plasma beta \citep{matsukiyo2014} is typically $\beta_{\rm PUI}=2\alpha M_A^2\sim 12.8$ in the outer heliosphere where the ratio of the PUI to solar wind plasma density $\alpha\sim 0.1$ and Alfv\'en Mach number $M_A\sim 8$.
Hence, the total plasma beta becomes much greater than $1$.
\citet{li2021} recently constructed a magnetic reconnection model that includes the effect of thermal pressure based on \citet{liu2017}, finding that the reconnection rate tends to decrease in a high-plasma beta environment.
This is a consequence of the heated plasma downstream of the reconnection region stagnating the outflow, which acts to decelerate the outflow and results in a low reconnection rate.
Therefore, this suggests that PUIs decrease the reconnection rate.

However, this expectation is based on an ideal-fluid perspective exclusively, and does not include some important PUI-induced physics, these being PUI-driven turbulence, heat conduction and viscosity.
The turbulence is driven by newly-born PUIs, which form initially a ring-beam distribution function that is unstable and resonant with parallel-propagating Alfv\'en waves \citep{isenberg2003}.
This PUI-driven turbulence has been observed as a spectral enhancement in the solar wind turbulence \citep{hollick2018} and is strong enough to heat the proton core component as a consequence of the dissipation of turbulence \citep{williams1995,matthaeus1999,isenberg2003}.
PUI-heat conduction and viscosity are the result of the first- and second-order anisotropies in the PUI velocity distribution function, respectively \citep{zank2014}.
\citet{mostafavi2017} shows that these effects efficiently smooth the HTS, and that the thickness of the smoothed transition region is consistent with Voyager 2 observations \citep{richardson2008}.
As we discuss in detail below, these effects have the possibility of increasing the reconnection rate.
Therefore, the simplified description of PUI-magnetic reconnection in the absence of the additional effects introduced by PUI-induced physics can lead to incorrect conclusions about PUI-mediated reconnection in the heliosphere.

In this Letter, we predict that PUI-induced turbulence, heat conduction, and viscosity can act preferentially to increase the reconnection rate at heliospheric current sheets in the outer heliosphere.
This Letter focuses on the region beyond the ionization cavity ($\sim10$ au) where PUIs are expected to play a dominant role in the thermodynamics of the solar wind and consider H$^+$ PUI.
Note that since there have been several reports of reconnection events in HCSs at $1$ au \citep{gosling2006, gosling2007}, it is likely that reconnection also occurs in HCSs in the outer heliosphere. 

\section{PUI-induced Turbulence, Heat Conduction, and Viscosity} \label{sec:style}

The possibility that turbulence increases the magnetic reconnection rate in space and astrophysical plasmas has been considered before \citep{matthaeus1985, lazarian1999}.
This can be interpreted as a consequence of decreasing the current sheet aspect ratio due to magnetic field wandering introduced by the turbulence.
According to \citet{higashimori2013}, in which they perform 2D Reynolds-averaged MHD simulations \citep{yokoi2011}, the condition for the boost in magnetic reconnection due to turbulence can be described in terms of the quantity $C_\tau=\sqrt{C_\beta}{\tau_{\rm turb}}/{\tau_A}$,
where $C_\beta$ is a turbulence model parameter ($C_\beta=0.3$)\citep{higashimori2013}, $\tau_{\rm turb}$ the turbulence time scale, and $\tau_A=\delta/v_A$ the Alfv\'en crossing time ($\delta$ is the half-thickness of a current sheet).
When $C_\tau$ falls into the range of $0.8\le C_\tau \le 1.6$ , the reconnection rate is dramatically increased \citep{higashimori2013}.
Using this criterion, we check whether PUI-driven turbulence causes explosive reconnection.

\begin{figure}[htbp]
  \centering
  \includegraphics[scale=1.5]{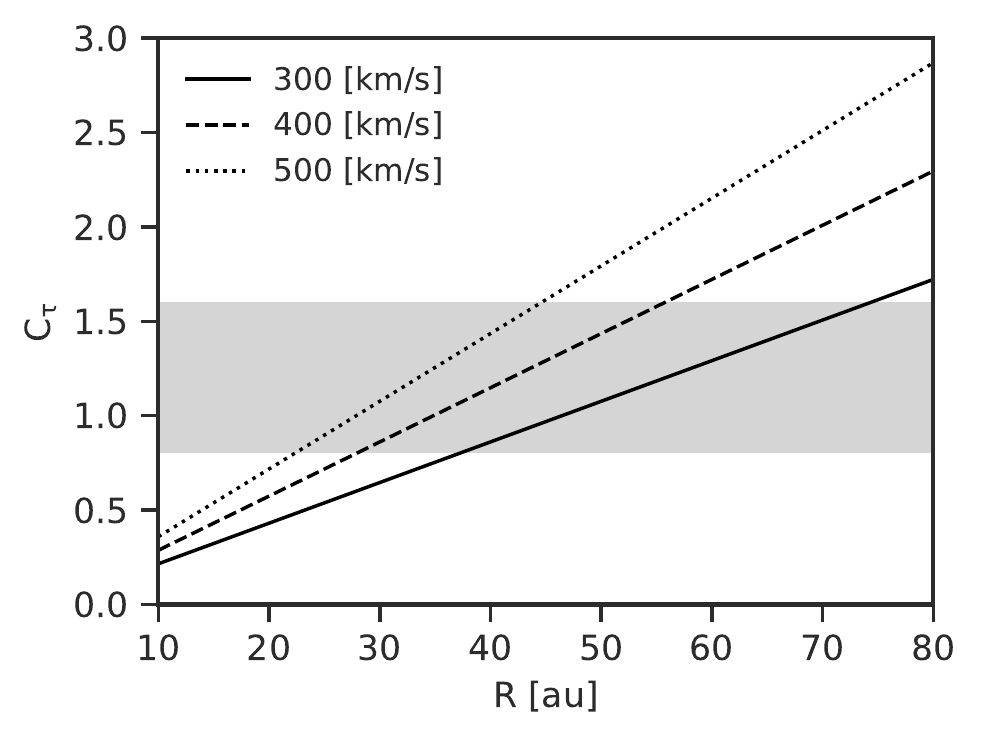}%
  \caption{Reconnection boost parameter, $C_\tau$, due to PUI-driven turbulence for several PUI characteristic speeds $c_p=300$, $400$, and $500$ km/s, which are equivalent to solar wind speed. The grey shadowed region corresponds to the condition of explosive magnetic reconnection \citep{higashimori2013}. \label{fig:Ctau}}
\end{figure}

PUI-driven turbulence provides a favorable mechanism to boost magnetic reconnection.
Assuming that the PUI-driven turbulence is highly Alfv\'enic \citep{zank2000}, we can approximate the turbulence time scale as $\tau_{\rm turb}\approx \lambda_{||}/v_A$, where $\lambda_{||}$ is the parallel turbulence length scale and defined by $\lambda_{||}=2\pi c_p/\Omega_c$ \citep{isenberg2005, oughton2011}.
Here,  $c_p$ is the PUI characteristic speed, which is equivalent to solar wind speed, $\Omega_c$ is the proton-cyclotron frequency, and we assume that the magnitude of the interplanetary magnetic field is described as $B=B_0(R_0/R)$ with $B_0=0.5$ nT at $R_0=10$ au.
The condition $C_\tau$ then results in $C_\tau=\sqrt{C_\beta}{\lambda_{||}}/{\delta}$.
The half-thickness of HCSs is assumed to be $\delta\sim10^5$ km \citep{liou2021}.
Fig. \ref{fig:Ctau} shows $C_\tau$ for several PUI characteristic speeds, $c_p=300, 400$, and $500$ km/s, which we assume a constant solar wind speed for simplicity.
The gray-shadowed region corresponds to the condition of explosive turbulent reconnection.
For instance, with a PUI characteristic speed of $400$ km/s, efficient turbulent reconnection can occur from $30$ to $55$ au. 
The slower ($300$ km/s) and faster ($500$ km/s) speed covers $20-40$ and $30-70$ au, respectively.
Therefore, this indicates that the PUI-driven turbulence can boost magnetic reconnection over a wide range of distances for several PUI characteristic (or solar wind) speeds.

\begin{figure}[htbp]
  \centering
  \includegraphics[scale=1.5]{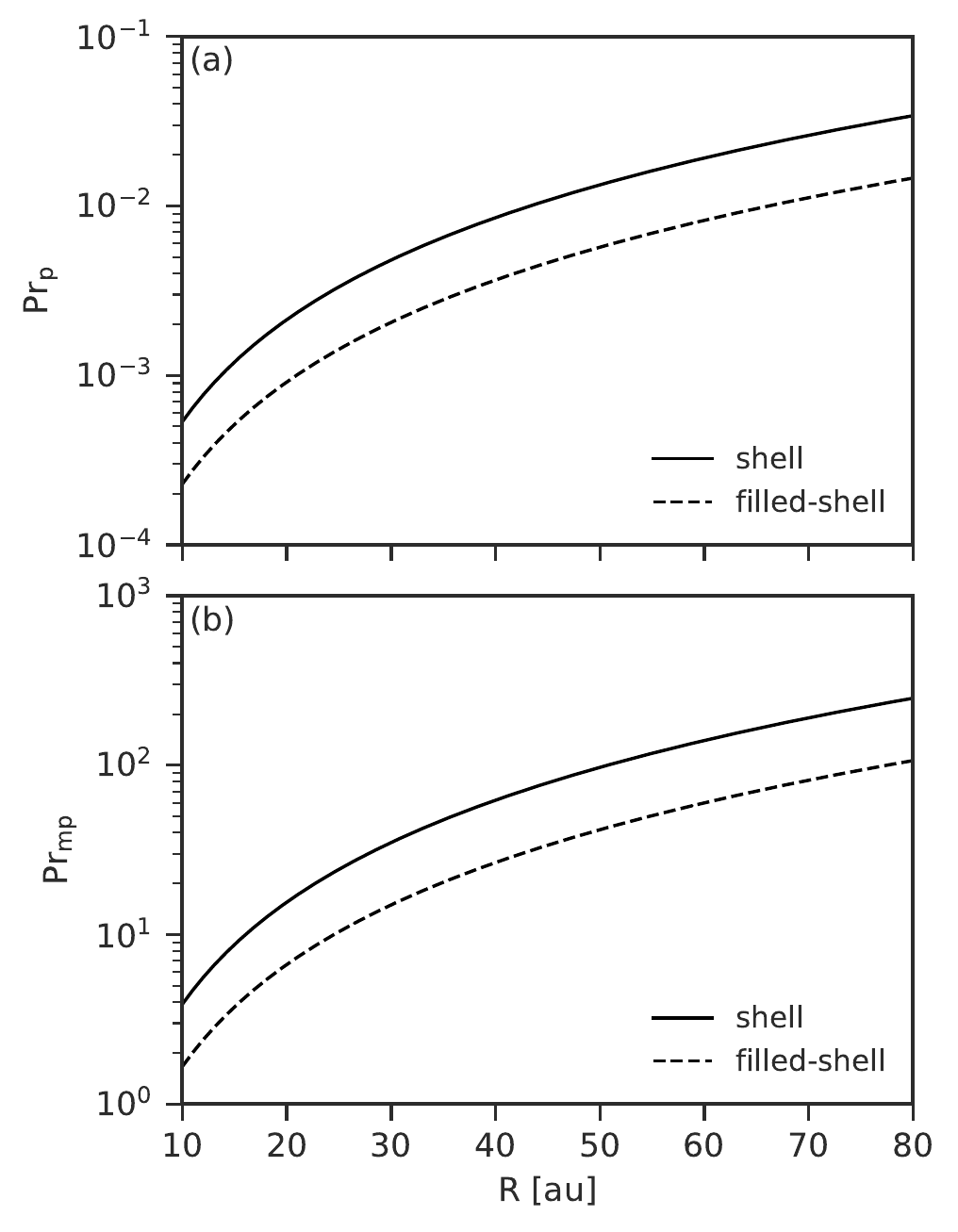}%
  \caption{(a) PUI-fluid Prandtl number $Pr_p=\nu_p/(m_pn_t\kappa_p)$. (b) PUI-magnetic Prandtl number $Pr_{mp}=\nu_p/(m_pn_t\eta)$. Solid and dashed lines correspond to shell and filled-shell distributions.\label{fig:Prandtl}}
\end{figure}

The PUI-induced heat conduction and viscosity yields the fluid and magnetic Prandtl numbers, ${Pr_p}=\nu_p/(m_pn_t\kappa_p)\ll 1$ and ${Pr_{mp}}=\nu_p/(m_pn_t\eta)\gg 1$ in the outer heliosphere.
Here, $m_p$ is the proton mass, $\nu_p$ the PUI-viscosity, $\kappa_p$ the PUI-heat conduction, and $\eta$ the magnetic resistivity in the solar wind.
We set the total plasma density to be $n_t=n_0(R_0/R)^2+n_p$ where a core-plasma density $n_0=0.1$ cm$^{-3}$ at $R_0=10$ au, and a constant PUI density $n_p\sim8\times10^{-4}$ cm$^{-3}$ \citep{mccomas2021}.
According to \citet{zank2014}, $\nu_p=P_p\tau_s/15$ and $\kappa_p=c_p^2\tau_s/3$, where $\tau_s$ is the scattering time of PUIs, and we set $\tau_s=\Omega_{c}^{-1}$.
We assume that the PUI pressure can be derived from either a shell ($P_p=m_pn_pc_p^2/3$) or filled-shell ($P_p=m_pn_pc_p^2/7$) distribution \citep{vasyliunas1976,zank2016}.
We use a Spitzer formalism for magnetic resistivity, $\eta=d_e^2\tau_e^{-1}$ \citep{spitzer1962} where $d_e$ is the electron inertial length, and $\tau_e$ the dissipation timescale.
Here, we assume $\tau_e=d_i/v_{te}$ based on the scale of wave-particle interaction \citep{verma1996} where $d_i$ is the ion (proton) inertial length, and $v_{te}$ the electron thermal speed, being set as a constant using a temperature of $10^4$ K \citep{mccomas2021}. 
Fig. \ref{fig:Prandtl} shows the PUI-fluid (top panel) and magnetic (bottom panel) Prandtl numbers for a shell (solid) and filled-shell (dashed) distributions.
Here, we use a constant PUI characteristic speed, $c_p=400$ km/s.
Both the fluid and magnetic Prandtl numbers follow $Pr_p, Pr_{mp}\propto R^2$.
We can see that the fluid (magnetic) Prandtl number is much less (larger) than $1$ in the outer heliosphere.
This indicates that the PUI-viscosity dominates the magnetic resistivity, and the PUI-heat conduction process is much faster than the viscous time scale. 
Note that the fluid Prandtl number does not depend on $c_p$.

These Prandtl numbers match a condition that corresponds to a boost in magnetic reconnection.
\citet{minoshima2016} perform 2D single-fluid MHD simulations that include heat conduction and viscosity effects, and find that a small fluid- and large magnetic-Prandtl number ($Pr<1$ and $Pr_m>1$) preferentially boosts magnetic reconnection.
Their interpretation is that having the viscosity stronger than the resistivity $(Pr_m>1)$ results in a broader vortex layer than the current sheet layer, and the quadrupolar vortex excited in the vortex layer efficiently carries the upstream magnetic flux towards the reconnection region and the reconnection rate increases.
The heat conduction plays a role in sustaining a current layer narrower than the vortex layer by convecting heating energy generated by viscous dissipation away from the reconnection site. 
Since the PUI-heat conduction and viscosity precisely satisfy the condition ($Pr_{p}\ll 1$ and $Pr_{mp}\gg 1$), we conclude that PUIs can effectively accelerate magnetic reconnection.

PUI-heat conduction introduces a further very interesting possibility, and this is to relax the condition of the \citet{li2021} reconnection model.
A primary reason for the decrease of the reconnection rate in a high-beta plasma is that the heated plasma downstream acts to decelerate the outflow.
Consequently, if the PUI-heat conduction efficiently removes the thermal (heated) energy along the magnetic field, this can lead to a relaxation of the \citet{li2021} model, with the result that the reconnection rate may increase. 
A similar effect has been found in single-fluid MHD simulations with heat conduction \citep{chen1999}.

\section{Conclusion} \label{sec:floats}

In this Letter, we have presented the possibility that PUI-induced turbulence, heat conduction, and viscosity preferentially boost magnetic reconnection at HCSs in the outer heliosphere.
This is contrary to the simplified expectation from a single-fluid point of view that PUIs provide a high-plasma beta condition and decrease the reconnection rate.
It, therefore, suggests that we must take into account the PUI-induced physics to understand magnetic reconnection in the presence of PUIs.
Since PUIs introduce the possibility of boosting magnetic reconnection, this indicates that PUIs can be accelerated efficiently in PUI-mediated magnetic reconnection.
This may be of particular interest to the acceleration of the anomalous cosmic ray component in the vicinity of the HCS and just downstream of the HTS \citep{zank2015, zhao2019}.
Future work will examine the mediation of magnetic reconnection due to PUIs using nonlinear simulations.
We plan to use multi-fluid MHD simulations that incorporate a turbulence model \citep{zank2012}, PUI-induced heat conduction and viscosity effects \citep{zank2014}. 
In the MHD simulation model, the PUI and thermal plasma components are treated separately.

\begin{acknowledgments}
     We acknowledge the partial support of an NSF EPSCoR RII
     Track-1 Cooperative Agreement OIA-2148653, partial
     support from NASA awards 80NSSC20K1783 and 80NSSC23K0415, partial
     support from a NASA IMAP sub-award under NASA contract
     80GSFC19C0027, and a NASA Heliospheric DRIVE Center award SHIELD 80NSSC22M0164.
\end{acknowledgments}


\bibliography{sample631}{}
\bibliographystyle{aasjournal}



\end{document}